\documentclass[aps, prl, preprint]{revtex4-1}
\usepackage{amsmath}
\usepackage{graphicx}
\usepackage{caption}
\usepackage{subcaption}
\usepackage{natbib}
\begin{document}
\title{Coherent and Incoherent Thermal Transport in Nanomeshes}
\author{Navaneetha K. Ravichandran}
\author{Austin J. Minnich}
\email{aminnich@caltech.edu}
\affiliation{Division of Engineering and Applied Science,\\ California Institute of Technology, Pasadena, California 91125, USA}
\date{\today}
\begin{abstract} 
Coherent thermal transport in nanopatterned structures is a topic of considerable interest, but whether it occurs in certain structures remains unclear due to poor understanding of which phonons conduct heat. Here, we perform the first fully three-dimensional, frequency-dependent simulations of thermal transport in nanomeshes by solving the Boltzmann transport equation with a novel, efficient Monte Carlo method. From the spectral information in our simulations, we show that thermal transport in nanostructures that can be created with available lithographic techniques is dominated by incoherent boundary scattering at room temperature. Our result provides important insights into the conditions required for coherent thermal transport to occur in artificial structures.
\end{abstract}
\maketitle
Heat conduction in solids at length scales comparable to phonon mean free paths (MFPs) and wavelengths is a topic of considerable interest~\cite{Cahill:2003aa, Zebarjadi:2012aa, Heremans:2013aa}. Recently, nanostructured materials such as nanowires~\cite{Boukai:2008aa, Hochbaum:2008aa}, superlattices~\cite{Chowdhury:2009aa}, nanocomposites~\cite{Poudel:2008aa, Mehta:2012aa, Ma:2013aa}, and all-scale nanostructured materials~\cite{Biswas:2012aa} have demonstrated strongly reduced thermal conductivities compared to their parent bulk materials. Many of these nanostructured semiconductor materials show promise for thermoelectric energy conversion~\cite{Boukai:2008aa, Hochbaum:2008aa, Chowdhury:2009aa, Poudel:2008aa, Mehta:2012aa, Ma:2013aa, Biswas:2012aa, Biswas:2011aa}.

Coherent thermal transport, in which the phonon dispersion is modified by the coherent interference of thermal phonons in an artificial periodic material, is an active area of research~\cite{Luckyanova:2012aa, Sadhu:2012aa, Yu:2010aa, Maldovan:2013aa, Maldovan:2013ab, Ravichandran:2014aa}. Unlike classical boundary scattering, which only decreases the relaxation times of phonons, coherent effects can also alter the group velocity and density of states by zone folding of phonons, which was originally observed at specific frequencies using Raman scattering~\cite{Colvard:1980aa}. A number of recent works have studied coherent transport both theoretically~\cite{Maldovan:2013aa, Maldovan:2013ab, Balandin:1998aa, Yang:2003aa} and experimentally. Several recent experiments have reported that coherent effects can affect thermal transport in superlattices and nanomeshes~\cite{Luckyanova:2012aa, Yu:2010aa, Ravichandran:2014aa}. In particular, exceptionally low thermal conductivities were reported in silicon nanomeshes (NMs)~\cite{Yu:2010aa, Hopkins:2011aa} which consist of periodic pores in a thin membrane. However, attributing these low thermal conductivity measurements unambiguously to coherent effects is difficult because boundary scattering can also reduce the thermal conductivity of the NMs. Therefore, despite these experimental observations, the conditions under which coherent thermal transport can occur remain unclear.

Computational studies have attempted to provide insight into coherent transport in artificial periodic nanostructures. \textsl{Hao et al}~\cite{Hao:2009aa} used two dimensional Monte Carlo simulations of phonon transport to predict a reduction in thermal conductivity of porous silicon with aligned pores. \textsl{Jain et al}~\cite{Jain:2013aa} used a mean free path sampling algorithm to study phonon transport in NM-like structures with features larger than 100 nm, concluding that coherent effects are unlikely to be the origin of the low thermal conductivity in the structures of Ref.~\cite{Hopkins:2011aa}. \textsl{He et al}~\cite{He:2011aa} investigated NM-like structures with similar surface area to volume ratio as in Ref.~\cite{Yu:2010aa} but the simulated structures were much smaller than those studied experimentally. \textsl{Dechaumphai et al}~\cite{Dechaumphai:2012aa} used a partially coherent model, in which phonons with MFP longer than the NM neck size were assumed to be coherent, to explain the observations of Ref.~\cite{Yu:2010aa} but boundary scattering could not be rigorously treated in their analysis. Thus, due to several simplifications and approximations used in these studies, the questions of which phonons are responsible for heat conduction in complex structures like NMs and under what conditions coherent transport can occur in these structures remain unanswered.

To address this issue, we present the first fully three-dimensional simulations of thermal transport in NMs using efficient numerical solutions of the frequency-dependent Boltzmann transport equation (BTE). Using the spectral information in our simulations, we find that coherent thermal transport is likely to occur at room temperature only in structures with nanometer critical dimensions and atomic level roughness, and that boundary scattering dominates transport in structures that can be created lithographically. Our work provides important insights into the conditions in which coherent thermal transport can occur in artificial structures.

We begin by describing our simulation approach. To gain spectral insights into the heat conduction in nanostructures we must solve the frequency-dependent BTE, given by~\cite{Majumdar:1993aa},
\begin{equation}
\frac{\partial e_{\omega}}{\partial t} + \underline{v}\cdot\nabla _re_\omega = -\frac{e_\omega - e^0_\omega}{\tau _\omega}
\end{equation}
where $e_\omega$ is the desired distribution function, $\omega$ is the angular frequency, $e^0_\omega$ is the equilibrium distribution function, $\underline{v}$ is the phonon group velocity and $\tau _\omega$ is the frequency-dependent relaxation time. 

The phonon BTE has been solved for simple nanostructure geometries using several techniques such as the discrete ordinate method~\cite{Yang:2004aa}, Monte Carlo simulation method~\cite{Jeng:2008aa}, the finite volume method~\cite{Narumanchi:2005aa} and the coupled ordinates method~\cite{Loy:2012aa}. However, using these methods to accurately simulate transport in the large 3D geometry of the NM is extremely challenging due to computational requirements or due to the use of simplifying approximations that may not be applicable.

We overcome this challenge by solving the BTE with an efficient variance-reduced Monte Carlo algorithm, achieving orders of magnitude reduction in computational cost compared to other deterministic or stochastic solvers~\cite{Peraud:2011aa, Peraud:2012aa}. Briefly, this technique solves the linearized energy-based BTE by stochastically simulating the emission, advection and scattering of phonon bundles, each representing a fixed deviational energy from an equilibrium Bose-Einstein distribution. The variance of the simulation is reduced compared to traditional MC by properly incorporating deterministic information from the known equilibrium distribution in a control variates approach. This algorithm enables the first simulations of thermal transport directly in the complex 3D geometry of the NM.

To implement the algorithm, the phonon dispersion is divided into 1000 frequency bins, and the phonon bundles are emitted into the simulation domain according to the appropriate distribution as described in Ref.~\cite{Peraud:2011aa}. Since the scattering operator is linearized in this approach, the phonon bundles are advected and scattered sequentially and completely independently of each other. 

\begin{figure*}[ht]
\begin{minipage}[b]{0.45\linewidth}
\includegraphics[height=0.25\textheight]{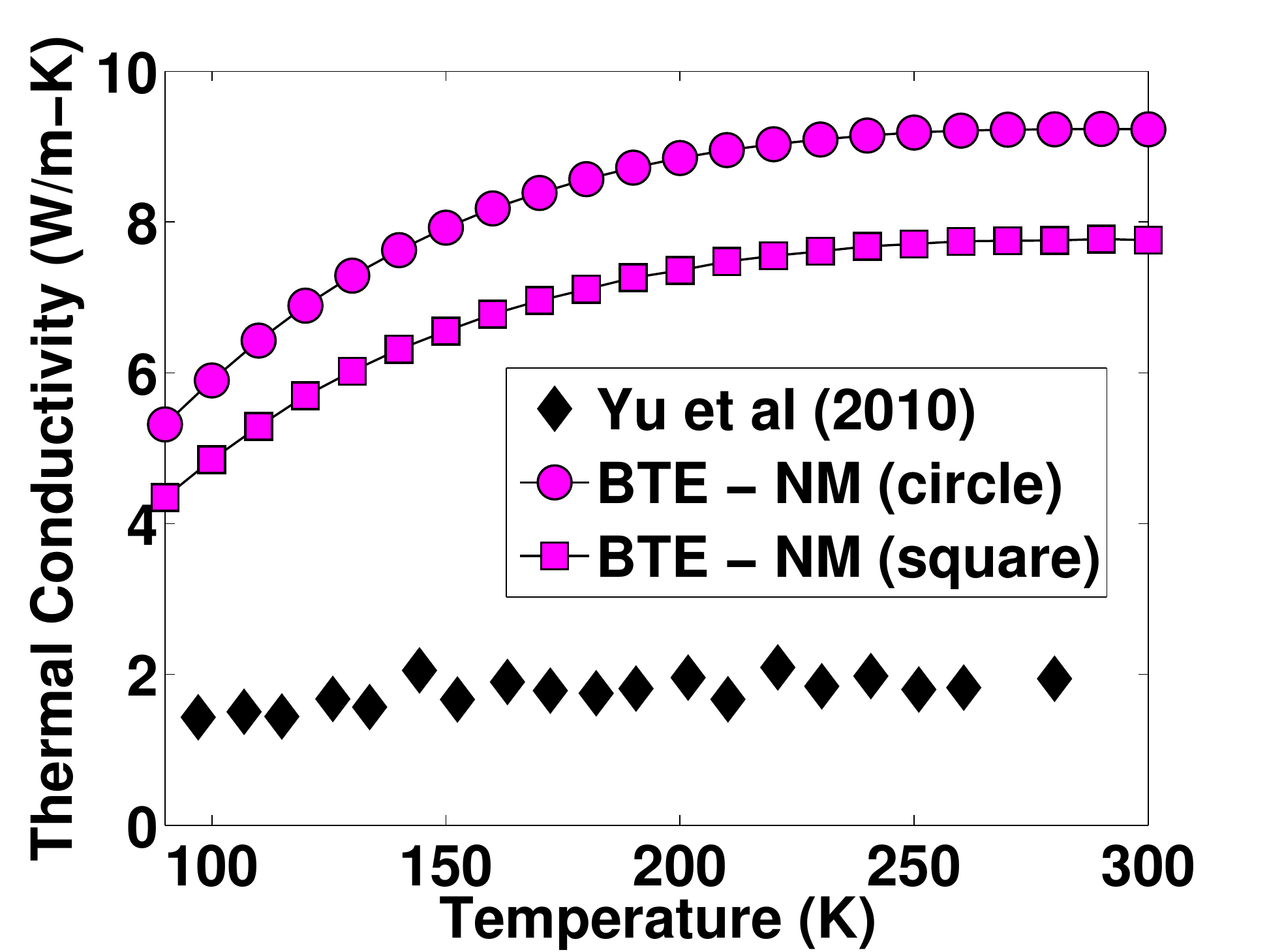}
\caption{Thermal conductivity of the NM as a function of temperature. The thermal conductivity of the NM reported in Ref.~\cite{Yu:2010aa} (black diamonds) is significantly lower than our simulation result with square and circular pore geometries.}
\label{motivation}
\end{minipage}
\hspace{0.5cm}
\begin{minipage}[b]{0.45\linewidth}
\includegraphics*[trim=0mm 0mm 10mm 0mm, height=0.25\textheight]{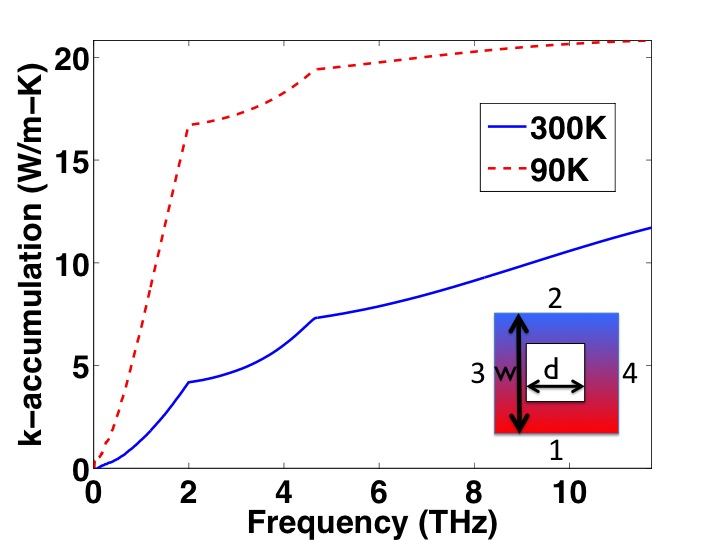}
\caption{Thermal conductivity accumulation versus phonon frequency for a NM that reflects phonons with frequency less than 2 THz specularly and the rest diffusely. Even under these conservative assumptions, the reported measurements cannot be explained even by completely neglecting the contribution of these low frequency phonons that could undergo coherent interference.} \label{SpectralK}
\end{minipage}
\end{figure*}

To compute the thermal conductivity of the NM, we simulate the thermal transport in a single periodic unit cell of the NM using periodic heat flux boundary conditions~\cite{Hao:2009aa} as indicated in the inset of Fig.~\ref{SpectralK} (boundaries 1 and 2). The other two periodic walls of the NM (boundaries 3 and 4) are modeled as specularly reflecting boundaries since the unit cell of the NM is symmetric about its center. The top and bottom boundaries in the out-of-plane direction and the walls of the NM pores are modeled as diffusely reflecting mirrors. The thermal conductivity of the NM is computed by adding up the contribution of the trajectory of each phonon bundle to the overall heat flux. We terminate the propagation of phonons after $10$ internal scattering events as the change in thermal conductivity of the NM is less than $0.5\%$ between the tenth and the twentieth internal scattering event. 

We use an isotropic Si dispersion along the [100] direction and phonon relaxation times from Ref.~\cite{Minnich:2011aa}. To validate our simulation, we calculate the thermal conductivity of an unpatterned silicon thin film doped with Boron. We find that we can explain the reported measurements assuming the boundaries scatter phonons diffusely and using the impurity scattering rate of the form $\tau^{-1}_{\mathrm{Imp}} = 2\times 10^{-44}\ \omega ^4\ s^{-1}$, where $\omega$ is the angular frequency of phonons. For the NM simulations, we consider both circular and square pores as the shape of the NM pore is somewhere in between. Electron-phonon scattering is expected to be negligible at the temperatures considered~\cite{Asheghi:2002aa} and is not included. 

We begin our analysis by computing the thermal conductivity of a NM structure. To facilitate comparisons with experiment, we simulate the same structure in Ref.~\cite{Yu:2010aa} with a periodicity w $=34$ nm, a pore width or diameter d $=11$ nm and an out-of-plane thickness t $=22$ nm. Since all the physical walls of the NM are modeled as diffusely reflecting mirrors, our MC simulations yield the Casimir limit for the thermal conductivity of the NM, which is the theoretical lower limit for the thermal conductivity of the NM with phonons following the unmodified bulk dispersion. It is evident from our simulation results (Fig.~\ref{motivation}) that the experimentally measured thermal conductivity of the NM is considerably lower than the Casimir limit.

We now examine whether coherent transport can explain this exceptionally low thermal conductivity. According to \textsl{Jain et al}~\cite{Jain:2013aa}, for coherent effects to occur in periodic nanostructures, long wavelength phonons, which are more likely to scatter specularly from a rough boundary and retain their phase, should conduct most of the heat. At present, the minimum phonon wavelength that can scatter specularly from a surface with a given roughness remains unclear, with estimates ranging from 0.64 THz~\cite{Maznev:2013aa} to 2 THz~\cite{Luckyanova:2012aa}. From these experimental observations, we can infer that coherent effects could affect phonons below $2$ THz, while the remaining part of the phonon spectrum will still follow the bulk material dispersion and lifetimes.

Conservatively, let us suppose that phonons with frequency below $2$ THz may be able to follow the new dispersion corresponding to the phononic crystal. When we assume that the boundaries of the NM reflect these low frequency phonons specularly so that their contribution to the relative fraction of heat transport is maximized, we obtain a thermal conductivity of $11.71$ W/m-K at 300K and $20.8$ W/m-K at 90K as shown in Fig.~\ref{SpectralK}. Even if we assume that phonons below $2$ THz behave coherently and completely remove their contribution to heat transport, the thermal conductivity of the NM reduces to $7.5$ W/m-K at 300K and $4.12$ W/m-K at 90K, which is still significantly higher the measured values of $1.95$ W/m-K at 300K and $1.3$ W/m-K at 90K in \textsl{Yu et al}'s experiments. A similar conclusion is reached if all phonons are scattered diffusely. Therefore, even under the most conservative assumptions, those modes that have the possibility to undergo coherent interference do not carry sufficient heat to explain the measurements.

We now use our simulations to identify the mechanism responsible for the experimentally observed reduction in thermal conductivity. Although \textsl{Yu et al}~\cite{Yu:2010aa} assumed that the NM was completely composed of silicon, in other experiments~\cite{Tang:2010aa} a thin amorphous oxide layer of about $2-3$ nm thickness is clearly visible using transmission electron microscopy, even though the samples were etched in HF vapor. Other studies have reported that surface damage can result from the reactive ion etching (RIE) process~\cite{Song:2004aa} used to create the pores in the NM.

The presence of such a disordered layer substantially affects the phonon transport within the NM. A phonon incident on the disordered layer from silicon has a probability to be backscattered at the interface before reaching the solid-air interface of the NM pores. Even if the phonon penetrates into the disordered layer, it will get scattered nearly immediately due to its short MFP in the disordered layer. Therefore, this disordered layer effectively increases the size of the pore and reduces the cross-sectional area available for heat conduction. 

This increased pore size has an important effect on the interpretation of experimental measurements. In the experiments of \textsl{Yu et al}~\cite{Yu:2010aa}, the thermal conductance of the NM was measured, and the thermal conductivity was calculated by assuming that heat effectively flows through channels between arrays of pores in the NM. If the effective size of the NM pores is larger than assumed, then the width of the heat transport channels is reduced, thereby increasing thermal conductivity for a given thermal conductance of the NM. Therefore, in order to interpret the experimental measurements in Ref.~\cite{Yu:2010aa} and compare with our simulations, the thermal conductivity of the NM has to be scaled by the ratio of the channel areas without and with the defective layer.

The large pores also lead to additional phonon boundary scattering due to increased surface area of the pores. To account for this effect in our MC simulations, we model the Si-disordered layer interface as a diffusely reflecting mirror. This is a reasonable approximation considering that the microscopic details of phonon scattering at interfaces is poorly understood~\cite{Lampin:2012aa}. The effective pore size is increased by an amount comparable to the thickness of the oxide layer as observed in TEM, which is around $2-3$ nm~\cite{Tang:2010aa}. We also include the disordered layer on the top and bottom boundaries of the NM as they were subjected to many of the same etching processes as the pores.
\begin{figure*}[!ht]
\centering
\includegraphics*[scale=0.45]{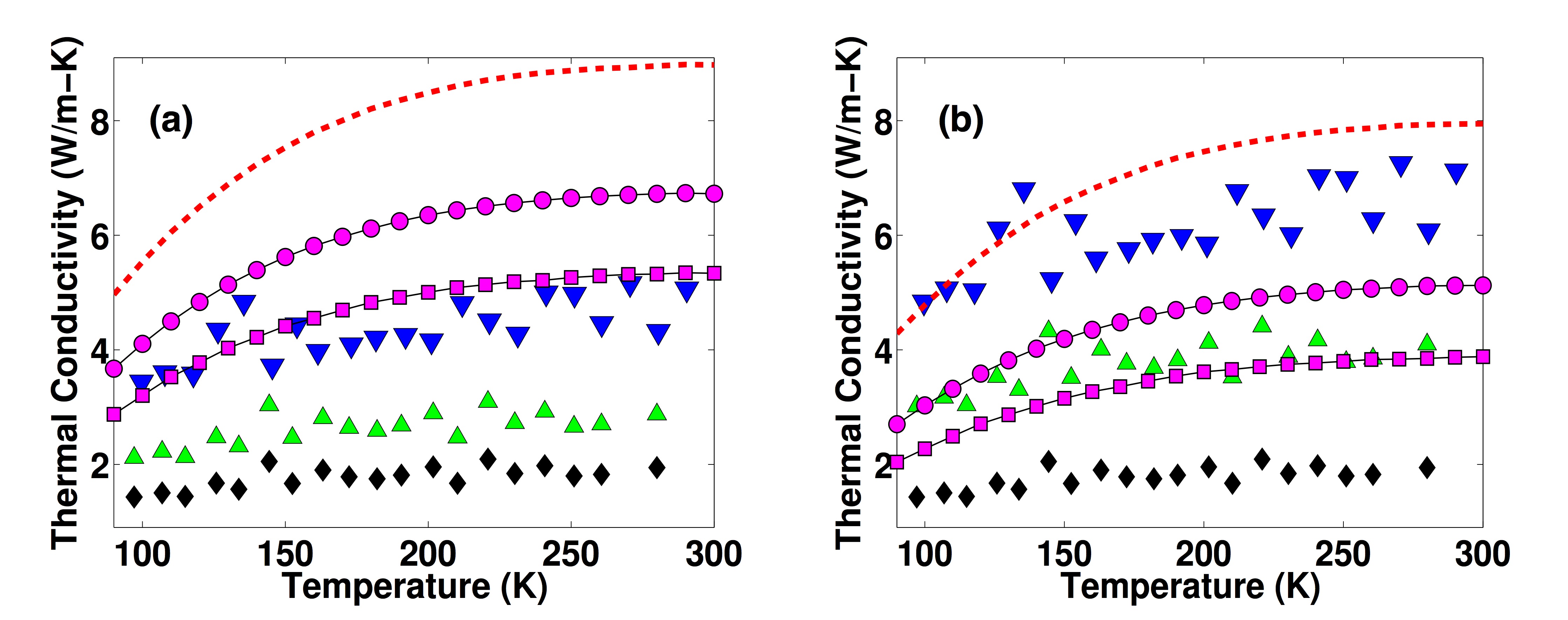}
\caption{Thermal conductivity as a function of temperature for different nanostructures from experiments in Ref.~\cite{Yu:2010aa} and our simulations for (a) $2$ nm disordered layer thickness and (b) $3.5$ nm disordered layer thickness. The disordered layer is added to both the NWA and the NM in our simulations. In these two figures, the red dashed line, pink circles, and pink squares are the MC solutions for the NWA, NM with circular holes, and NM with square holes, respectively. The black diamonds, green triangles and blue inverted triangles represent the reported thermal conductivity for the NM, the recalculated thermal conductivity for the NM, and the recalculated thermal conductivity for the NWA, respectively, from Ref.~\cite{Yu:2010aa}.} \label{final_kplot_2nm_3_5nm}
\end{figure*}

We now examine if the increase in the effective pore size can explain the observed reduction in the thermal conductivity of the NM. Figure~\ref{final_kplot_2nm_3_5nm}(a) shows that our simulations predict a considerable reduction in thermal conductivity of the NM for a disordered layer thickness of just $2$ nm, compared to the case without a disordered layer (Fig.~\ref{motivation}). As shown in Fig.~\ref{final_kplot_2nm_3_5nm}(b), we are able to explain the experimental observations with a $3.5$ nm thick disordered layer. \textsl{Yu et al}~\cite{Yu:2010aa} also reported the thermal conductivity for another NM with a larger pore (d = $16$ nm) at lower temperatures. By following the same simulation procedure, we are able to explain the measurements for this NM using a $2$ nm thick disordered layer.

Our simulations can also explain the difference in thermal conductivity between the NM and NWA. In \textsl{Yu et al}'s experiments~\cite{Yu:2010aa}, the reduction in thermal conductivity of the NM was associated with coherent effects primarily because of the lower thermal conductivity of the NM compared to the NWA even though boundary scattering considerations would predict the opposite trend. However, our simulations predict that the thermal conductivity of the NM is consistently lower than that of the NWA without considering any coherent effects.

This difference in thermal conductivity can be explained by backscattering of phonons at the walls of the NM pores ~\cite{Moore:2008aa}. In the NWA, all of the domain walls are aligned parallel to the direction of the thermal gradient and $50\%$ of the incident phonons are backscattered on average. The walls of the NM pores aligned along the temperature gradient also backscatter half of the incident phonons. However, the walls of the NM pores that are not aligned with the temperature gradient backscatter more than half of the incident phonons. Since backscattering reduces the contribution of the phonon to thermal transport, the overall thermal conductivity of the NM is reduced compared to the NWA.
\begin{figure}[!ht]
\begin{center}
\includegraphics*[scale = 0.4]{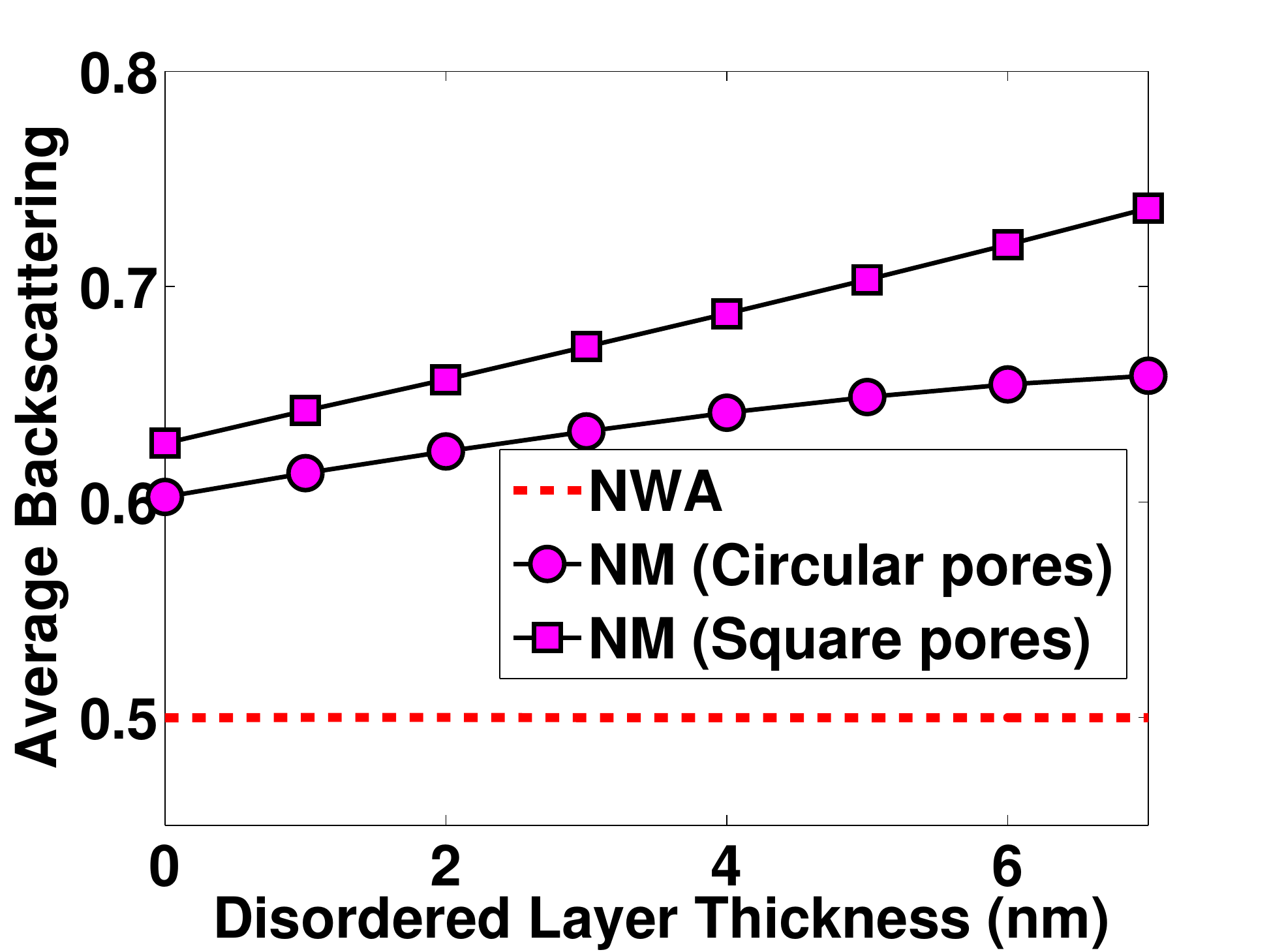}
\end{center}
\caption{Fraction of backscattered phonons for the NWA and the NM with circular and square holes for different disordered layer thicknesses.} ~\label{backscattering_vs_OX}
\end{figure}
Figure~\ref{backscattering_vs_OX} shows the fraction of backscattered phonons in the NWA and the NM averaged over all frequencies. We consider a phonon to be backscattered if it returns to the same wall from which it was emitted. For Fig.~\ref{backscattering_vs_OX}, to isolate the effect of phonon backscattering from the effects of difference in the size of the NWA and the NM, we simulate a NM and NWA with the same effective transport channel area. For the NM, we use a periodicity w $=34$ nm, pore size d $=12$ nm and thickness t $=22$ nm so that it has an effective transport channel area of $22\times 22$ nm$^2$. For the NWA, the cross-sectional area is $22\times 22$ nm$^2$. To isolate the effect of the geometry, we compute the backscattered fraction from those phonons that do not scatter internally in the domain. As expected for the NWA, $50\%$ of the phonons are backscattered. For the NM with circular and square pores, the fraction of backscattered phonons is $20-40\%$ higher than that of the NWA for a range of disordered layer thickness values used in our simulations. Therefore, the difference in thermal conductivity between the NM and NWA can be attributed to the larger fraction of backscattered phonons in the NM along with the smaller transport channel area of the actual NM.

We now examine the conditions under which coherent transport could occur in an artificial structure at room temperature. From the spectral information in our simulations, we find that most of the heat is carried by phonons with frequencies around 5 THz at room temperature, corresponding to a wavelength of about $1-2$ nm in Si. Therefore, a secondary periodicity on the order of this value is necessary for coherent effects to affect thermal transport in the NM. Further, the surface roughness of an artificial structure must be less than a few $\AA$ to preserve the phase of the scattered phonons. Such fine spatial resolution and atomic scale roughness is difficult to obtain using lithographic techniques, but could be met in superlattices with epitaxial interfaces~\cite{Ravichandran:2014aa}. In lithographically patterned structures, coherent thermal transport is likely to play a role only at very low temperatures where the dominant thermal wavelength substantially exceeds the surface roughness amplitude.

In conclusion, we have performed the first fully three-dimensional simulations of thermal transport in nanomeshes using efficient numerical solutions of the frequency dependent BTE. From the spectral information in our simulations, we find that incoherent boundary scattering dominates thermal transport in lithographically patterned structures, and that structures with nanometer critical dimensions and atomic level roughness are required for coherent thermal transport to occur at room temperature. Our results provide important insights into the conditions under which coherent thermal transport can occur in artificial structures.

This work is part of the 'Light-Material Interactions in Energy Conversion' Energy Frontier Research Center funded by the U.S. Department of Energy, Office of Science, Office of Basic Energy Sciences under Award Number DE-SC0001293.
\bibliographystyle{apsrev4-1}
\bibliography{Heath_Coherence}
\end{document}